\documentclass[preprint2]{aastex631}

\shorttitle{Identifying catastrophic outliers in the COSMOS field}
\graphicspath{{./}{figures/}}

\usepackage{paralist}
\usepackage{amsmath}
\usepackage{verbatim}
\usepackage{graphicx}
\usepackage{rotating}
\usepackage{float}
\usepackage{todonotes}
\usepackage{xcolor, soul}

\usepackage{orcidlink}

\begin{document}
\title{Identifying catastrophic outlier photometric redshift estimates in the COSMOS field with machine learning methods}


\author[0000-0001-9066-0552]{Mitchell T. Dennis}
\affiliation{Institute for Astronomy \\
University of Hawai`i at M\=anoa \\
2680 Woodlawn Drive \\
Honolulu, HI 96822, USA}

\author[0009-0008-7427-4617]{Esther M. Hu}
\affiliation{Institute for Astronomy \\
University of Hawai`i at M\=anoa \\
2680 Woodlawn Drive \\
Honolulu, HI 96822, USA}

\author[0000-0002-6319-1575]{Lennox L. Cowie}
\affiliation{Institute for Astronomy \\
University of Hawai`i at M\=anoa \\
2680 Woodlawn Drive \\
Honolulu, HI 96822, USA}

\begin{abstract}
    We present the result of two binary classifier ensembled neural networks to identify catastrophic outliers for photo-z estimates within the COSMOS field utilizing only 8 and 5 photometric band passes, respectively. Our neural networks can correctly classify 55.6\% and 33.3\% of the true positives with few to no false positives. These methods can be used to reduce the errors caused by the errors in redshift estimates, particularly at high redshift. When applied to a larger data set with only photometric data available, our 8 band pass network increased the number of objects with a photo-z greater than 5 from 0.1\% to 1.6\%, and our 5 band pass network increased the number of objects with a photo-z greater than 5 from 0.2\% to 1.8\%.
\end{abstract}

\section{Introduction}
\label{sec:intro}
\par Current and upcoming large-scale sky surveys (e.g. HEROES \citep{2023arXiv230211581T}, COSMOS \citep{2022ApJS..258...11W, 2016ApJS..224...24L, Ilbert_2008, Capak_2007}, GOODS \citep{2003mglh.conf..324D}, LSST \citep{2019ApJ...873..111I}, SDSS \citep{2017AJ....154...28B}) have or will take observations for millions of galaxies within a limited number of photometric bands. These measurements will provide researchers with redshifts for weak lensing, enabling them to study the large scale structure of the universe, and to draw conclusions about the observed galaxies and determine which ones warrant follow-up investigations using spectroscopy. Redshift is among the most important criteria, and therefore it is crucial that calculations of the redshift based on photometric data (e.g. using algorithms such as LePhare \citep{1999MNRAS.310..540A, 2006A&A...457..841I}, EAZY \citep{2008ApJ...686.1503B}, SPIDERz \citep{2017A&A...600A.113J}, and \citep{2012MNRAS.420.1217D} from SDSS) be accurate. 
\par While these algorithms have been largely successful, they are not perfect. The choice of method can influence the accuracy of the result. The standard approach is to fit observations to a representative set of templates (template fitting) and minimize $\chi^2$ (e.g. \citep{2000A&A...363..476B}). This method does not require spectroscopic information and therefore is not limited by the need for spectroscopic observations. Other methods utilize training algorithms with known spectroscopic redshifts to map photometric data to redshifts. Both of these techniques are limited by the data that is used to create them (i.e. the template set or the training data) and can be unreliable. Other techniques such as using Bayesian statistics (e.g. \citep{2000ApJ...536..571B}) or Monte Carlo methods (e.g. \citep{2001AJ....122.2205R}) offer their own advantages and disadvantages. All photometric redshifts rely on the features shown on the observed spectral energy distribution (SED). For objects whose SEDs have less distinct features or whose distinct features (e.g. the Lyman $\alpha$ break) have been substantially shifted into wavelengths that were not covered or easily observed, acquiring accurate photometric redshifts is even more challenging.
\par Although photometric redshift techniques have improved over time, photometric redshift catalogues continue to contain small populations of galaxies with redshifts that are substantially under- or over-predicted. Many different works (e.g. \citet{2022ApJ...928....6S, 2020PASP..132b4501J}) have established different definitions for outliers and catastrophic outliers (COs). In this work we investigate a specific subset of COs, deliberately focusing on COs that populate the high redshift universe. For our study, this subset includes all objects with a photometric redshift (photo-z) $0 \leq \textrm{photo-z} \leq 1.5$ and the spectroscopic redshift (spec-z) satisfies $1.8 \leq \textrm{spec-z} \leq 7.0$. The data in this range also satisfies the condition where the difference between the spectroscopic redshift (spec-z) and the photo-z is greater than 1.5. All other objects are designated as non-catastrophic outliers (NCOs). The range we have chosen is an important component of the analysis of the COSMOS dataset (e.g \citep{2014frdy.conf..463F}). Objects with less distinct features in their Spectral Energy Distributions (SEDs) and those with distinct features that have been substantially redshifted cause misclassification of objects by the photometric redshift algorithm. Additionally, high redshift work that currently focuses on Lyman $\alpha$ emitters and rarer objects of other sorts, may be particularly affected by the use of SED fitting. Selectively targeting known difficult regions of parameter space with data-driven corrections will improve the quality of photometric redshifts.
\par As with all incorrect predictions, COs negatively impact the scientific products of these catalogues, and therefore developing algorithms and methods for identifying these outliers is crucial to achieving the scientific goals for these surveys. Rather than creating an entirely new redshift algorithm, this work focuses on positively flagging these objects for potential correction and different weighting, or even removal from large galaxy studies. Previous works in this area include \citet{2017A&A...600A.113J, 2019A&A...621A..26P, 2022arXiv220104391M} using various machine learning (ML) methods including both neural networks and support vector machines (see \citep{2021FrASS...8...70B} for a review).

\par The data selection criteria are given in \S \ref{sec:data}, a discussion on the results of ML testing is in \S \ref{sec:training}, an application of the network is discussed in \S \ref{sec:applying}, we investigate the source of COs in \S \ref{sec:cause}, a discussion of our work in \S \ref{sec:discuss} and our conclusions in \S \ref{sec:conclusions}.

\section{Selecting the Data}
\label{sec:data}
\par This work utilized the photo-zs provided in the COSMOS2020 photo-z catalogue \citep{2022ApJS..258...11W} hereafter COSMOS2020. These data include RA, DEC, a set of photo-z solutions and their corresponding chi-squared solutions, as well as various magnitudes and their errors. The photo-zs used were calculated using the \textit{Le Phare} template fitting method introduced in \citep{Arnouts_1999} with subsequent improvements in \citep{Ilbert_2006}. Of the photometric data, only the u, B, V, g, r, i, z, and K magnitudes were used. We utilize only these few band passes as they provide a relatively large number of data points that have non-null values. Null values, even those documented with a representative number (e.g. 99) are problematic for machine learning as they introduce spurious relationships and so must be discarded. To address this we limit our magnitude criterion to $-80 < \textrm{ABS Mag} < 80$ where ``ABS Mag" is any of our magnitudes. Any object either completely missing photometric data or whose data indicates an observation but does not have a detection in one of these bands was dropped from the sample. Moreover, in some instances, a specific band pass was missing but could be substituted with a similar one (e.g., the Hyper Suprime-Cam I Band magnitude was missing, but within the same dataset, there was a Suprime-Cam I Band magnitude). In these cases the substitution was made in an attempt to maximise the size of our dataset. Due to the similarity of the band passes, we do not expect this to substantially affect the machine learning.

\par
\begin{figure}
    \centering
    \includegraphics[scale=0.5]{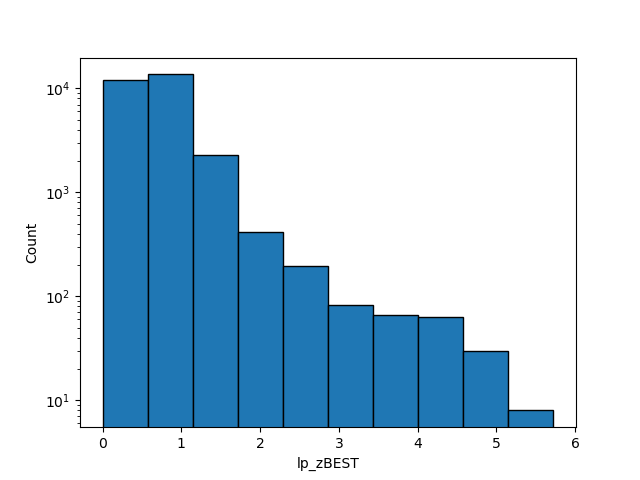}
    \caption{A histogram of roughly 29,000 photometric redshifts for the training data that is fed into the ML algorithm, presented on a logarithmic scale on the y-axis. The plot reveals that the majority of objects have a redshift less than 2. Within the training data, where both photo-z and spec-z are available, only a small percentage of objects have a spec-z greater than 2 (2.9\%), greater than 3 (0.94\%),  greater than 4 (0.45\%), and greater than 5 (0.10\%).}
    \label{fig:histTraining}
\end{figure}

\begin{figure}
    \centering
    \includegraphics[scale=0.5]{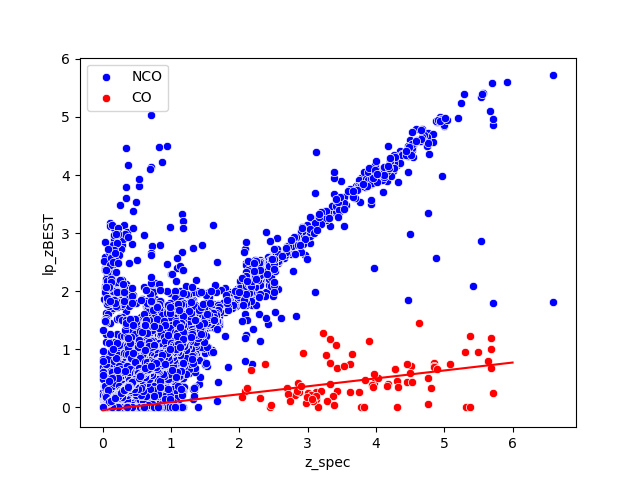}
    \caption{A plot of the training data set comparing photometric redshifts to spectroscopic redshifts. The objects highlighted in red constitute a small but crucial subset for understanding the high redshift universe. This subset includes all objects with a photometric redshift (photo-z) less than 2.0 and a difference between photo-z and spectroscopic redshift (spec-z) exceeding 1.5. Visual inspection suggests that this is a reasonable starting point for our cut. These objects represent a significant percentage of cases where spec-z$ > 2$. \S \ref{sec:data} provides a more detailed discussion on the imbalanced nature of the dataset as a whole, which may be harder to discern from this figure alone. The trend-line, along with a $68\%$ confidence interval, is utilized in \S \ref{sec:training} to correct flagged outliers. In \S \ref{sec:data}, we explore how this simple correction affects the overall redshift distribution in the entire dataset. The trend-line has an R$^2$ value of 0.14, indicating that a linear relationship explains $14\%$ of the variation between photo-z and spec-z. The p-value for the R value is small $0.0003$, with a standard error of $0.036$. This p-value suggests that it is highly unlikely the correlation is due to chance.}
    \label{fig:scatter}
\end{figure}

 Spectroscopic data were taken from the DEIMOS 10k spectroscopic sample \citep{Hasinger_2018}, zCOSMOS spectroscopic data \citet{Pozzetti_2010}, and the current publicly available spectroscopic data from the COSMOS field (Salvato, M. 2021, private communications). For these data, we restricted our study to the redshift measurements with a quality flag of greater than 1. Paraphrasing \citep{2007ApJS..172...70L}, of the classes we kept in our sample, class 4 are considered ``completely secure redshift" with ``unambiguous multiple spectral features;" class 3 are also "very secure" but with a possibility of error, class 2 are still secure, but there is a significant risk of error. These are in contrast to class 9 where, objects for which there is a single unambiguous emission feature, but it cannot be completely determined which feature it is between H$\alpha$, O$[II]$, or Ly$\alpha$, and so the most likely case was chosen, class 1, which is ``an informed guess", and class 0 which indicates no redshift could be calculated. Although class 2 has a significant risk of error due, we still include them because there are studies conducted which either fail to explicitly list the quality flag selection criteria (e.g. \citep{2024arXiv240702973S, 2024arXiv240619437T, 2024ApJ...967...60J}) and others that include quality flag 2 (e.g. \citep{2024ApJ...970...50C}), typically with an acknowledgement that the results for these objects are tentative. Our analysis focuses exclusively on objects with quality flag between 2 and 4,  excluding quality flags greater than 10, which correspond to AGN with the 11, 12, 13, and 14 corresponding to flags 1, 2, 3, and 4 respectively. The quality flag and photometric data restrictions yielded 86 COs. Of these remaining 86, 19 have a quality flag of 4, 28 have a quality flag of 3, 38 have a quality flag of 2 (22.1\%, 32.5\%, and 45.4\% respectively). A histogram of the photo-z distribution for our training data are shown in Figure \ref{fig:histTraining}, and a scatter plot identifying the specific subset of COs that this work focused is shown in Figure \ref{fig:scatter}. Furthermore, Figure \ref{fig:scatter} shows that COs account for 14.5\% of the $\textrm{spec-zs} > 2$, 24.4\% of the $\textrm{spec-zs} > 3$, 24.6\% of the $\textrm{spec-zs} > 4$, and 37.9\% of the $\textrm{spec-zs} > 5$. The fraction of sources with reported spectroscopic redshift after we made our photometric cuts with spec-z greater than 2, 3, 4, and 5 is 2.1\%, 0.94\%, 0.45\%, and 0.10\%.  A visual inspection of Figure \ref{fig:scatter} shows a relatively clean grouping where we begin our cut that all appear to follow a similar trend. This is in contrast to other works which typically define COs as any object where the absolute distance between the photo-z and the spec-z is greater than 1 \citep{2022ApJ...928....6S, 2021PASP..133d4504W, 2018AJ....155....1G, 2010MNRAS.401.1399B}. Because we used the same data that produces the photo-z (and thus also produced this trend), we reasoned there was relationship in parameter space that could be exploited by machine learning to identify these values. The trend-line is shown on the chart and has an R$^2$ value of 0.14, with the p-value of 0.0003 being nearly zero, i.e. it is unlikely that this trend is due to chance. There is not enough statistical confidence in the fit of this line to recover accurate photo-zs, but this toy model can give some idea of the number of very high redshift galaxies that may be currently under reported.

\par Our ML network employs ensemble learning, which involves using numerous ML networks in parallel. Their results are combined to form a single output to maximize accuracy. Therefore, training, testing, and validation datasets were created for the ML ensemble as a whole. These splits were broken down as 80\% training, 10\% testing, and 10\% validation. These are subsequently referred to as the ensemble data sets. We built 30 individual networks into our ensemble, and used the ensembled nature of our algorithm to account for the errors in the photometric data.

\par ML networks tend to perform better on data that has been normalized, and so we normalized the magnitude and error data between zero and one. Further, each entry in our catalogue includes error bars for each of the magnitudes and a confidence interval for the photo-zs. For an individual observation, we treated the errors in the magnitudes as standard deviations and randomly sampled Gaussian distributions centered on the ABS magnitude each of the photometric band passes using a standard deviation equivalent to the given error (in mag units) each respective band pass. This procedure was repeated for each individual network in the ensemble for the training data alone. Our final ensemble network had 30 individual constituents, and therefore 30 slightly altered versions of the training data were created. 

\par Our dataset contains 28,841 NCOs and 86 COs ($\approx0.30\%$ COs, a highly unbalanced dataset) as shown in \ref{fig:scatter}. ML networks perform best on balanced data sets \citep{2019SPIE11198E..13S} (e.g. datasets which have a 1:1 ratio between different classifications). Highly unbalanced datasets are a persistent problem in ML and have been addressed by developing tools to balance them. We balanced our data by applying the Synthetic Minority Oversampling Technique (SMOTE) \citep{2011arXiv1106.1813C} on the individual training data sets. Instead of randomly removing data (and subsequently losing potentially valuable information), SMOTE creates synthetic data points for the minority class(es). SMOTE randomly selects a member of the minority class, calculates the convex combination between this member and its neighbors also in the minority class, and randomly selects a point within the convex combination. This process is repeated until the data set is balanced. For our data this was only done for the training data only. Unbalanced datasets will continue to pose a persistent challenge, and further tools are being developed to address this problem.

\par To ensure a proportional split between training, testing, and validation sets, the two classes (COs and NCOs) were separated before being randomly divided into training, testing, and validation data sets for each class (yielding 6 data sets). The data sets were then combined with their corresponding counterparts into the training, testing, and validation data sets.

\section{Creating the Network}
\label{sec:training}
\subsection{Constructing the Network}
\par As discussed above, our neural networks are implemented in an ensemble, meaning the results of multiple (in our case 30) neural networks are aggregated together into a single binary output (CO or NO). We implemented our neural networks using TensorFlow and its companion package Keras \citep{tensorflow2015-whitepaper, chollet2015keras}. The networks were implemented using the standard Adam optimizer \citep{2014arXiv1412.6980K} with additional controls including an early stopping condition and a monitor that adjusts the learning rate. While each model is slightly different due to the random nature of each of the individual training data sets, the networks generally converge within 75 iterations (or epochs). The loss function used was binary crossentropy loss.

\par The hyperparameters used for the Adam optimizer were a learning rate of 0.001, $\beta_1 = 0.9$, and a $\beta_2 = 0.9$. The early stopping condition monitored the validation data loss metric (binary crossentropy) with a patience of 15 epochs, and the learning rate monitor also monitored the validation data loss metric and had a patience of 10 epochs with a factor of 0.3 reduction.

\par A table summarizing the architecture of the constituent networks within the larger ensemble is provided below in \ref{tbl:architecture}. The constituent networks each have 13 layers. There are 11 input columns (8 magnitudes and 3 photo-z values) for our first analysis, and 8 for our second analysis (5 magnitudes and 3 photo-z). These inputs are immediately passed to a batch normalization layer. During training, the batch normalization layer normalizes each batch's current inputs to have a center of 0 and a standard deviation of 1. During inference mode (used for  both validation and testing data), it uses a moving average and standard deviation of the batches seen during training. Following batch normalization are 5 dense-dropout layer combos which all have a rectified linear unit activation function across 256 neurons and a dropout rate of 50\%. The final layer is a single nueron with a sigmoid activation function that outputs a value between 0 and 1.

\begin{table}[htbp]
    \centering
    \begin{tabular}{|c|c|c|c|}
        \hline Number & Type & Neurons/ & Activation \\
        & & Rate & Function\\
        \hline 1 & Input & 11 & Linear \\
        \hline 2 & Batch Norm. & N/A & N/A \\
        \hline 3 & Dense & 256 & ReLU \\
        \hline 4 & Dropout & 0.5 & N/A \\
        \hline 5 & Dense & 256 & ReLU \\
        \hline 6 & Dropout & 0.5 & N/A \\
        \hline 7 & Dense & 256 & ReLU \\
        \hline 8 & Dropout & 0.5 & N/A \\
        \hline 9 & Dense & 256 & ReLU \\
        \hline 10 & Dropout & 0.5 & N/A \\
        \hline 11 & Dense & 256 & ReLU \\
        \hline 12 & Dropout & 0.5 & N/A \\
        \hline 13 & Dense & 1 & Sigmoid \\
        \hline
    \end{tabular}
    \caption{This table describes the architecture of the constituent neural networks that comprise the ensembled network. All densely (or fully) connected layers are given rectified linear unit activation functions except for the output neuron which is given a sigmoid activation function. All dropout layers are given a 50\% dropout rate and directly follow a dense layer (except the output layer). Only one batch normalization layer is used before the first fully connected layer directly following the input layer.}
    \label{tbl:architecture}
\end{table}

\par The results of the constituent neural networks were then ensembled together to create a final prediction. Each individual network made a prediction on the ensemble validation data. These predictions were evaluated using each individual network's custom threshold value and then rounded to 1 (CO) or 0 (NO). If a data point was evaluated by every individual network to be a CO, it would receive a combined score equal to the total number of networks.

\subsection{Training and Validating the Network - 8 bands}
The input columns for our training data were the absolute magnitude columns for the u, B, V, g, r, i, z, and K band passes (with slight randomness based on their errors described above), and LePhare photometric redshift columns including the upper and lower 68\% bounds. The spectroscopic redshift columns were not included in the ML training and were only used to classify our data as COs and NCOs.

\par Both the testing and validation data sets at each level are withheld from training the network. However, the validation data set was used to ``guide" the network towards a solution (used as an evaluation at every iteration of the training) whereas the testing dataset is a final test. After we completed the training, we evaluated the model using the validation dataset, aiming to adjust the ``threshold" for a positive indicator (in this case a CO). Generally, the threshold is set to 0.5 (or 50\%) as in \citep{2022ApJ...928....6S}. This means if an individual network outputs a value $\geq 0.5$ for given a piece of input data, the result would be considered a positive. Adjusting this threshold is also known as threshold tuning. The 50\% threshold led to a large number of false positives, so many that they outnumbered the true positives. We applied a threshold tuner (i.e. we adjusted our network so that a much higher result (e.g. $0.9$) would be needed for a prediction to be considered a positive) that found the threshold which minimized the false positive rate. This reduced the false positive rate to zero at the individual network level. Threshold tuning was also applied at the ensemble level to the combined score and further optimized to minimize false positives.

\par
\begin{figure}
    \centering
    \includegraphics[scale=0.5]{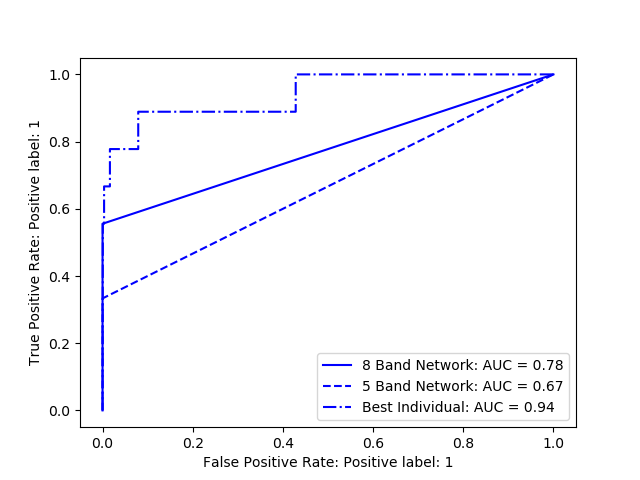}
    \label{fig:combo_roc}
    \caption{This plot shows the receiver operator characteristic (ROC) for the 8 band pass network, 5 band pass network, and the best individual performing model within the 8 band pass model. Their area under the curve (AUC) scores are 0.78, 0.67, and 0.94, respectively, out of a maximum possible 1.0. The scores of the combined ensembles are generally lower than their individual constituents (e.g. the 0.78 for the 8 band pass vs its individual 0.94) due to artificial restrictions placed on the ensembles themselves. Because the values coming into the ensemble scoring function have already been restricted to those constituent networks with a false positive rate of 0, only a false positive rate of 0 or 1 is possible. This has artificially reduced the the ensembles' ROC AUC scores.}
\end{figure}

 The ML algorithm correctly identified 55.6\% of COs. 20\% of these had a quality flag of 4, 20\% had a quality flag of 3, and the remaining 60\% had a quality flag of 2. Further, the number of false positives is zero which suggests all of the objects flagged by our network are COs that fall in the region of outliers shown in Figure \ref{fig:scatter}, and can be appropriately corrected. The final accuracy of the network is 99.86\%, outperforming the unbalanced nature of the data (remember the testing and validation data are as unbalanced as the original data). Our results are more accurately reported as a true positive rate of 55.6\% and a true negative rate of 100\%. Evaluation of the efficacy of machine learning classification algorithms are typically done using the receiver operator characteristic (ROC) and measuring the area under its curve. We supply ROC curves for both the best constituent model in our network and the ensemble as a whole in Figure \ref{fig:combo_roc}. The extremely high value of the ROC for our individual algorithms (on average near 0.9 and the best at 0.94) and our ensemble (0.78) indicates that our model performs well and only fails to identify all of the COs in an effort to eliminate false positives. The dramatic reduction in performance from the individual networks to the 0.78 of our ensemble is an artifact of how the ensemble was aggregated. Any model which failed produce a threshold that would yield 0 false positives was forced to zero. Therefore, all of the thresholds for the ensemble have a false positive rate of zero, artificially reducing the area under the curve of the graph compared to the value had thresholds been used in the individual models that allow for a non-zero false positive rate. Finally, the step-wise nature of the curves is caused by the low number of true positives (a feature of the unbalanced dataset) and is not by itself a cause for concern. Figure \ref{fig:combo_roc} also shows how much contamination is caused by false positives as the threshold is lowered to achieve a higher true positive rate.

\subsection{Training and Validating the Network - 5 bands}
\label{sec:test}
Upcoming photo-z challenges (e.g. LSST \citep{2018AJ....155....1G}) will unfortunately not include the K infrared band, which is crucial for distinguishing between different spectral breaks. Because of this, we repeated our entire analysis utilizing the same training data, network implementation and architecture to see if our success could be duplicated with fewer bands. Specifically, this new network only utilized the u, g, r, i, and z bands and attempted to classify the same COs and NCOs as above. All treatment of data, the number of networks in the ensemble, the randomization techniques, model hyperparamters, and threshold tuning were identical between the two networks. This process resulted in a network which was able to identify 33.3\% of COs. Of these, 33.3\% had a quality flag of 4, 33.3\% had a quality flag of 3, and 33.3\% had a quality flag of 2. The receiver operator characteristic curve for this new network is also provided in Figure \ref{fig:combo_roc}. The performance of this network is worse than the performance of the other network which is unsurprising given the further restrictions on the number of bands. Hereafter, the two networks will be called the 8 band network and the 5 band network.

\par Our results assume that spectroscopic redshifts are more accurate than their photometric counterparts (as is typical \citep{2016MNRAS.460.1371B}, but we acknowledge this as a limitation of our study in \S \ref{sec:discuss}). We can attempt to correct the identified outliers using the trend-ling from \ref{fig:scatter}. This trend-line is a function of variables photo-z and spec-z of the form $\textrm{photo-x} = m(\text{spec-z}) + b$. Our machine learning algorithm identifies objects that would fall in this region, and we have determined these objects follow this trend-line to a certain degree. 

\section{The cause of catastrophic outliers}
\label{sec:cause}
\par
\begin{figure}[htp]
    \includegraphics[scale=0.5]{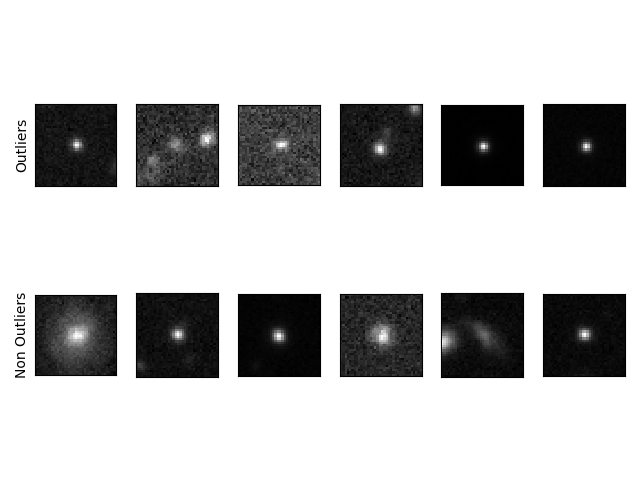}
    \caption{These images are a subsample of the examined postage-stamp images used to check the catalogue. There are 6 outliers and 6 nonoutliers. These postage-stamps are of the I band with a magnitude limit of $I < 25$, and were cut with a half-width of 3 arc seconds on each side. These images illustrate the challenge of discerning what might be causing outliers; there appear to be contaminating objects in both COs and NCOs (sub-figures 2, 4, and 11 numbered left to right, top to bottom}); and both NCOs and COs can also represent a range of compact and distributed objects.
    \label{fig:postage}
\end{figure}

 While this work primarily focuses on identifying COs using machine learning, we also attempt to discern the cause of COs, which remains unknown. During the preliminary stages of our study, we analysed both the photometric catalogue and spent time analysing the cutouts of the objects using data from the Hyper Suprime-Cam Subaru Strategic Program. Potential sources of error that could cause discrepancies between the photometric and spectroscopic redshift include the compaction of the object, object morphology, and contamination from nearby sources, particularly if those sources cannot be resolved individually. However, upon visual inspection of a large random sample of the NCOs, and all of the COs at different magnitude cuts, there was no notable difference between the fraction of objects that were faint sources, poorly resolved from nearby contaminants, or varying morphologies and compactness. We have included 12 postage stamps in Figure \ref{fig:postage} showing 6 sample objects of each class.

\par We further analysed the available tabular data (ABS Magnitudes) and from these also calculated several colors to evaluate the correlations between magnitudes, colors, and outlier classification. These analyses yielded little fruit in determining the source of the cause of COs. We have included charts showing the relationships between various colors and magnitudes in Figures \ref{fig:pairplot_magnitude} and \ref{fig:pairplot_color}. These figures demonstrate the relationship between outliers and non-outliers is not easily discerned in two dimensions and requires the multi-dimensional specialty of machine learning, specifically deep neural networks, which are better at extracting features than more standard regression analyses. We also produced a correlation matrix in Figure \ref{fig:heatmap} to calculate the linear correlations between the variables in our study and outlier status. Unfortunately, there were no strong correlations between variables and the outlier status that could identify the source of COs. Figures \ref{fig:heatmap}, \ref{fig:pairplot_magnitude}, \ref{fig:pairplot_color} are large and were therefore placed at the end of the text.

\par Source catalogue selection and quality flags are discussed in \S \ref{sec:data}, which also has plots detailing the distribution of photo-zs and the selected catastrophic outlier range and details on our sampling techniques. The results of these analyses are summarized in \S \ref{sec:discuss}, and a detailed correlation matrix can be found at the end of the text.

\section{Applying the Network}
\label{sec:applying}
After assessing our networks' accuracy, we applied our networks to the remaining photometric data which do not have spec-zs, but do have photometric data for the band passes we used to train them. For the 8 band network, the dataset initially contained 1,669,429 objects, of which 529,934 met all of the criterion in our 8 band passes and were within the range in photo-z of our training dataset $0 \leq \textrm{photo-z} \leq 6$. Of these, our 8 band pass classifier flagged 83,097 objects as COs after we discarded any of the outliers with a photo-z that were not in the range we defined as our outliers (photo-z [0, 1.5]). This is approximately $\approx 15.7\%$ of the data which met all of our criteria. However, the ML algorithm misses 45.4\% of COs, implying the true fraction of COs could be as high as 28.7\% in the most extreme case.

\par For the 5 band network, 587,395 objects met all of the criteria for our 5 band passes (u, g, r, i, and z). Of these, 59,544 were flagged as COs by our ML classifier after again discarding any that did not have an original photo-z between [0-1.5]. This is approximately $\approx 10.1\%$ of the data which met our criteria. However, this ML network misses approximately 66.6\% of COs implying the true number of COs could be $30.3\%$ in the most extreme case for these criteria. Both of these results are significantly higher than many other CO predictions based on similar data, motivating further studies to determine what the true fraction is and how to properly account for them.

\par Directly comparing the objects flagged by the 8 band and 5 band algorithms (83,097 and 59,544 respectively), 44,601 objects were flagged as outliers by both the 5 band algorithm and the 8 band algorithm; 14,472 objects were flagged by the 8 band pass algorithm and flagged by some of the constituent members of the 5 band pass algorithm but not enough to be considered an outlier by the 5 band pass algorithm as a whole; and 421 objects were flagged by the 5 band pass algorithm and flagged by some of the constituent members of the 8 band pass algorithm, but not enough to be considered an outlier by the 8 band pass algorithm as a whole. These numbers are not surprising, the 14,472 objects that were not flagged by the 5 band network may have been had the other available bands been provided, and the 421  additional outliers found by 5 band network are likely the product of the 5 band network considering roughly 60,000 more objects than the 8 band network.

\par
\begin{figure*}
    \centering
    \includegraphics[scale=1]{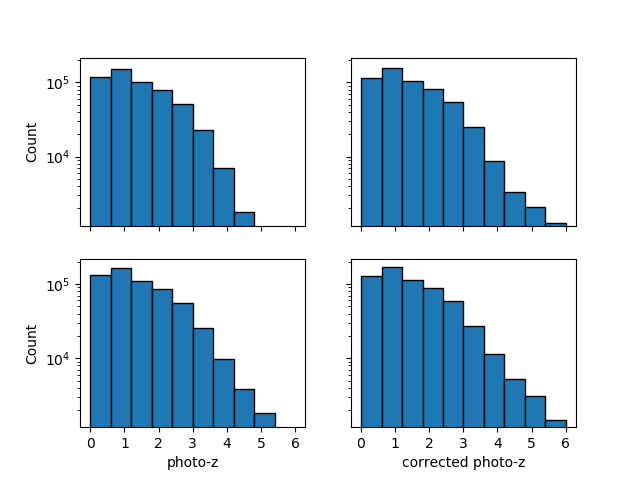}
    \caption{Four histograms comparing the original photometric redshifts from the data not used in training (no available spectroscopic redshifts) on the left with the corrected redshifts, obtained by applying the trend-line correction from Figure \ref{fig:scatter}, on the right for objects flagged by the 8 band pass network (top row) and the 5 band pass network (bottom row). The slight shift towards higher redshifts indicates a dramatic increase at high redshifts (Note the change in the base of the y-axis from left to right). For the 8 band pass network, the number of objects with a photo-z $ > 2$ increases from $\sim 21\%$ to $\sim 28\%$, photo-z $ > 3$ from $\sim 5\%$ to $\sim 10\%$, photo-z $ > 4$ from $\sim 0.7\%$ to $\sim 3.5\%$, and for photo-z $ > 5$ from $\sim 0.11\%$ to $\sim 1.6\%$. For the 5 band pass network, the number of objects with a photo-z $ > 2$ increases from $\sim 22\%$ to $\sim 27\%$, photo-z $ > 3$ from $\sim 6.4\%$ to $\sim 9.7\%$, photo-z $ > 4$ from $\sim 1.3\%$ to $\sim 3.6\%$, and for photo-z $ > 5$ from $\sim 0.2\%$ to $\sim 1.8\%$. For this analysis we only considered objects that, after the correction, fell within the original photo-z range. If all values of photo-z are considered, these ratios become slightly larger.}
    \label{fig:correction}
\end{figure*}

\begin{figure*}
    \centering
    \includegraphics[scale=0.3]{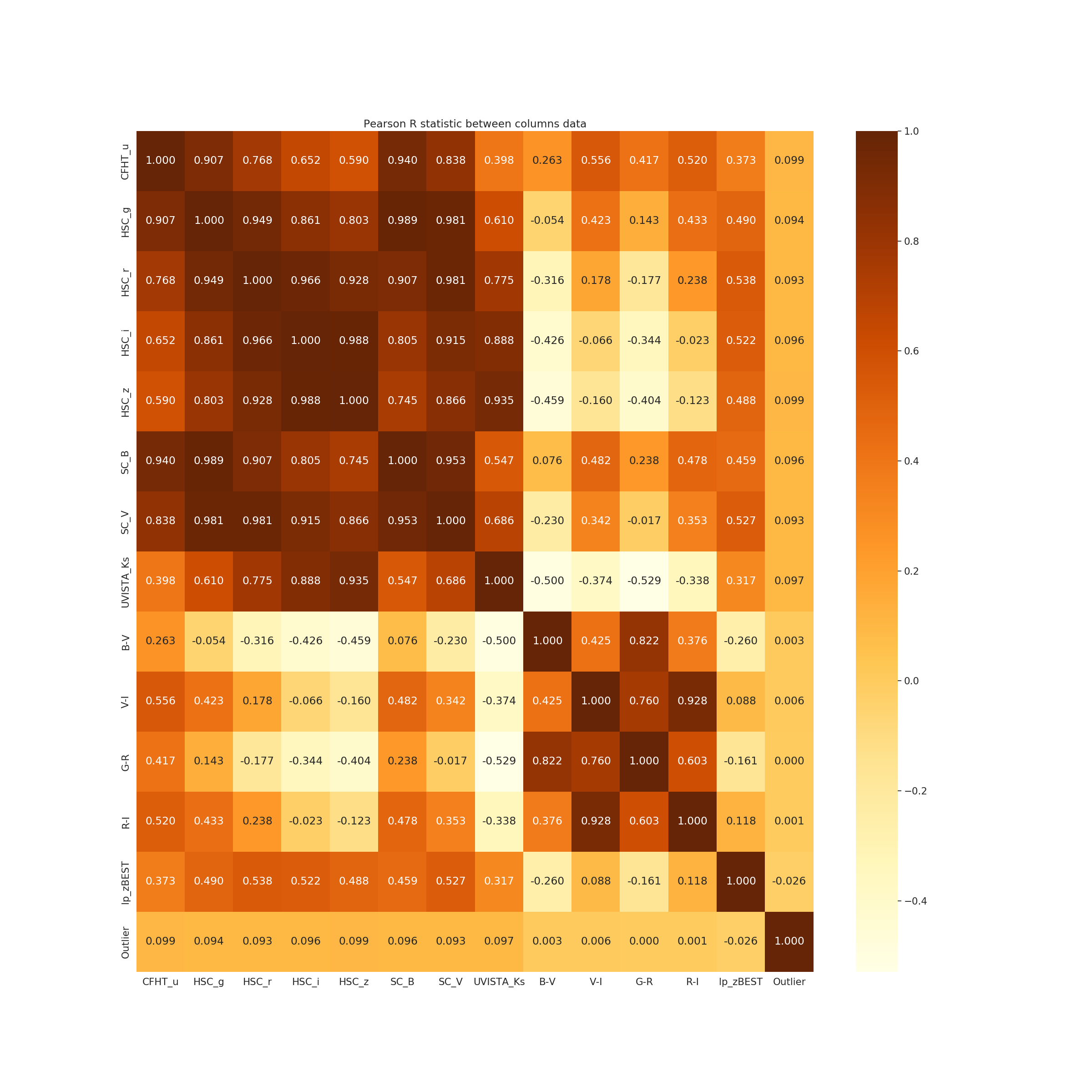}
    \caption{A correlation heatmap with data columns on the x and y axes and the Pearson R statistic labelled at each of the grid locations. The row and column labels correspond to the B, V, u, r, i, z, and Ks magnitudes, their associated errors, and the photo-z, and the upper and lower 68\% photo-z confidence intervals. While there are strong correlations between the various magnitude columns, and there are strong correlations between the photo-z and the 68\% confidence interval bounds, there are no strong correlations between any of the columns with our target variable (CO). Without strong correlations between the data columns, more advanced techniques such as machine learning are needed to accurately identify COs.}
    \label{fig:heatmap}
\end{figure*}

\begin{figure*}
    \centering
    \includegraphics[scale=0.3]{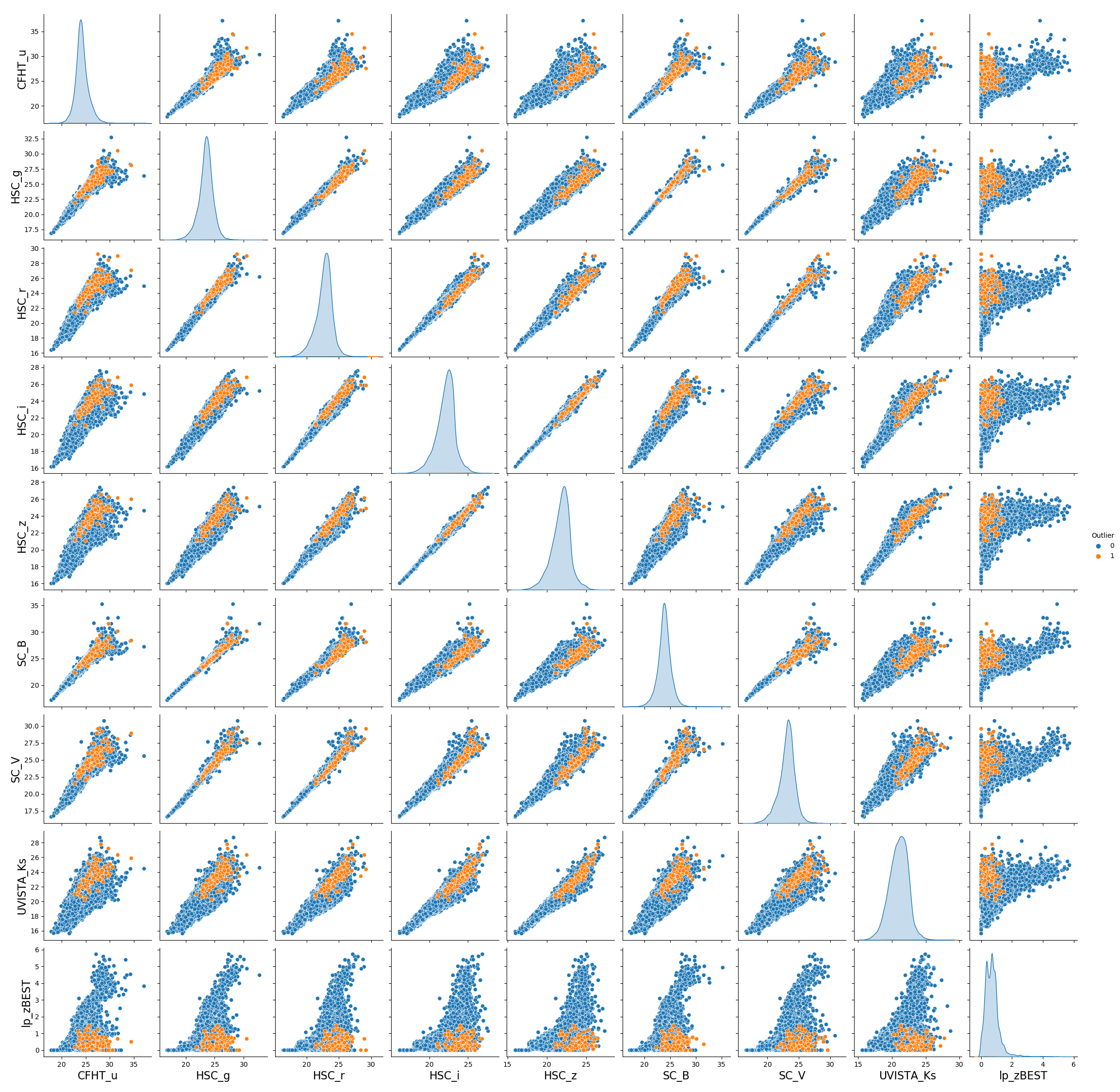}
    \caption{A series of scatter plots and Kernel Density Estimate (KDE) plots pairing each of the filter magnitudes together. The plots along the diagonal contain kernel density estimate (KDE) plots for each respective filter (and the photometric redshift). The scatter plots contain data corresponding to the x and y labels of each row or column. The upper triangle is simply a reflection of the lower triangle. These charts show almost no meaningful distinction between COs and NCOs in the magnitudes alone.}
    \label{fig:pairplot_magnitude}
\end{figure*}

\begin{figure*}
    \centering
    \includegraphics[scale=0.4]{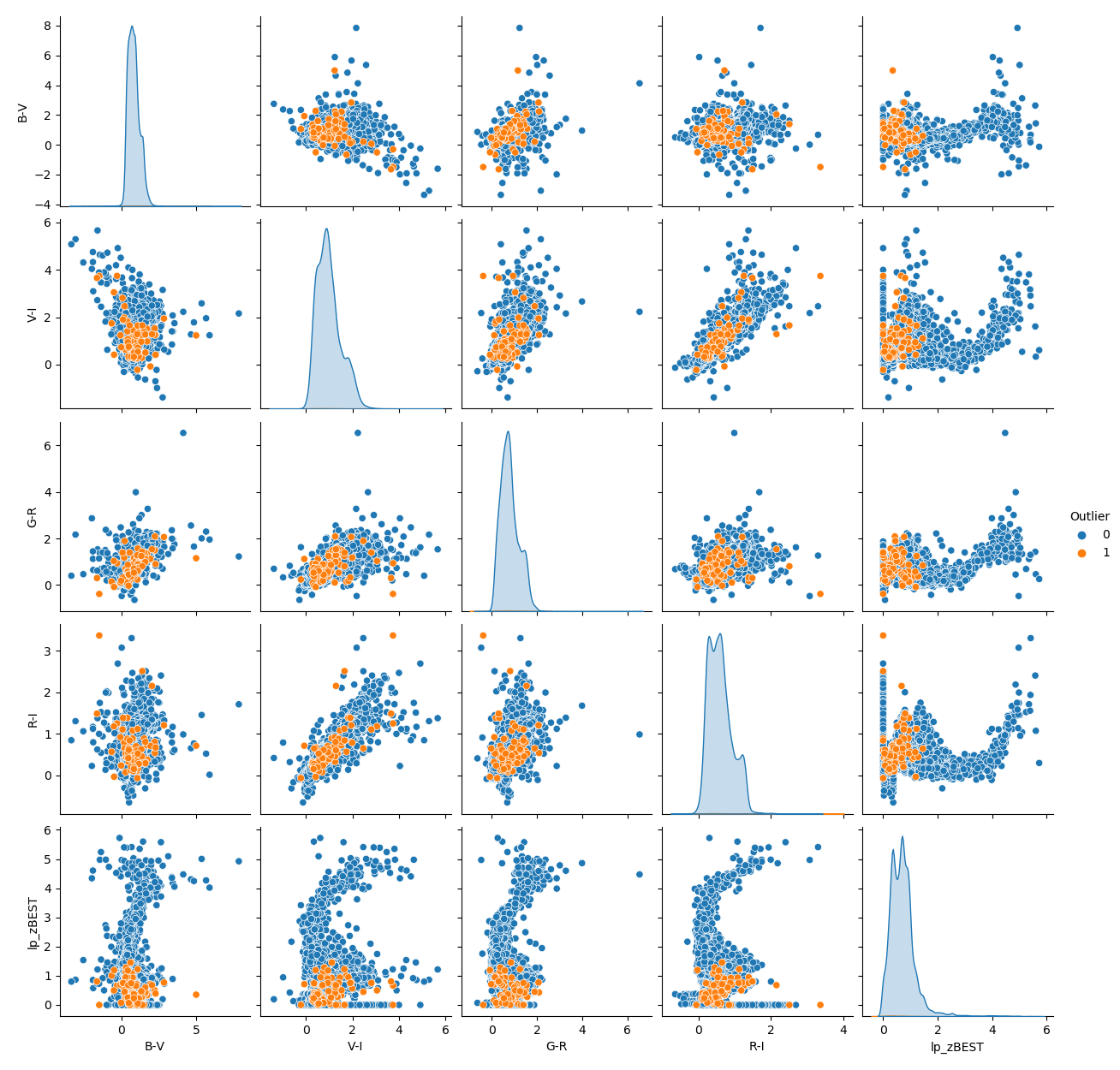}
    \caption{The same as Figure \ref{fig:pairplot_magnitude} but with four common colors instead of the magnitudes.}
    \label{fig:pairplot_color}
\end{figure*}

 Figure \ref{fig:correction} presents a set of histograms depicting the original photo-zs and ``corrected" photo-zs for both the 8 band pass network (top row) and the 5 band pass network (bottom row). The photo-z of the objects flagged by our 8 band network as COs can be adjusted using the inverse of the trend-line shown in Figure \ref{fig:scatter} (now of the form $\textrm{spec-z} = \frac{\textrm{photo-z} - b}{m}$ as a transformation to bring the photo-z from a CO to a value more closely resembling what the \textrm{spec-z} would be. This work does not claim definitive accuracy for these values, but it does provide a rough estimate of where a CO's true redshift is, again under the assumption the spectroscopic redshift values are more reliable. We caution readers that this trendline is a toy model, and that these values should not be used as photometric redshift. As previously stated, the linear model is not able to obtain highly accurate photo-z measurements, but instead can provide a rough picture of how many very high photo-zs are being under reported. 

\par For the 8 band pass network, the histograms in Figure \ref{fig:correction} illustrate an increase in the ratio of objects with a photo-z greater than 2 from 20.8\% to 28.0\%, an increase in objects with a photo-z greater than 3 from 5.4\% to 9.9\% , an increase in objects with a photo-z greater than 4 from 0.67\% to 3.5\%, and an increase in objects with a photo-z greater than 5 from 0.11\% to 1.6\%. The numbers for the 5 band pass histograms are as follows. The ratio of objects with a photo-z greater than 2 increases from 22.3\% to 26.4\%, greater than 3 increases from 6.4\% to 8.8\%, greater than 4 increases from 1.3\% to 3.6\%, and greater than 5 increases from 0.24\% to 1.8\%.

\par These numbers utilize only the data from the 529,934 and 587,395 objects that met our selection criteria, and we could only use our toy model on objects that our models flagged. Because our classifiers are not perfect, there may be more objects within these two sets that have low reported photo-zs, increasing these ratios. Conversely, assuming there are no COs in the remaining data, these changes could be reduced by as much as a factor of 3.1 or 2.8 respectively, assuming the distribution of photo-zs remains the same.

\par These results are a substantial modification to the COSMOS results and are explored further in \S \ref{sec:discuss}.

\section{Discussion}
\label{sec:discuss}
\subsection{Limitations of our study}
\par The first limitation of this study is our assumption that spectroscopic redshifts are consistently more reliable than their photometric counterparts. We aimed to address this by only using the spectroscopic data that had higher quality flags, but the fact remains that some of these spectroscopic redshifts may be incorrect and therefore our results overstated. The results of our testing data indicate half of the outliers may correspond to objects that would have a spectroscopic quality flag of 2 (the other half having a quality flag of 4). Although there is no way to filter these from our predictions, we can conservatively estimate that up to half of our identified COs in the prediction data may ultimately have less reliable spectroscopic redshifts.

\par Another limitation of our work is the very small number of true positives in our dataset. The number of NCOs is relatively large, enabling the ML to clearly label it in parameter space. However, the small number of COs makes the challenge of carefully identifying the true boundary of the CO space even harder. Because of this it is possible that the ML is predicting false positives in the photometric data even when the testing data returns a false positive rate of zero. This would also result in an overstatement of our results.

\par Additionally, our study focused on data with a detection in all of the available band passes. However, this eliminated a substantial fraction of the total data ($\approx$ 68\% for the 8 band pass network and $\approx$ 65\% for the 5 band pass network). Assuming the most conservative case (there are no COs in this data), this would reduce the number of COs detected to roughly 4.9\% for the 8 band pass network and 3.5\% for the 5 band pass network or 8.9\% and 10.7\% when accounting for the missed objects, respectively. If these are further combined with our first limitation, then we can conservatively estimate that the lowest rates of COs from our two results are $\approx$ 3.6\% and $\approx$ 7.1\% (only allowing for quality flags of 3 or 4), a 55.4\% and reduction from 8.9\% and a 33.3\% reduction from 10.7\%) for the 8 band and 5 band networks, respectively. However, considering that the other half of the objects have fewer detected band passes, and given the typical assumption that spectroscopic redshifts are more reliable than photometric redshifts, this most conservative estimate would seem unlikely.

\par Finally, it is important to note that while our results indicate an increase in the number of objects recorded at redshift greater than 2, other work focusing on the inverse of our area of interest (high photo-z but low spec-z) will almost certainly decrease the number of objects recorded at redshift greater than 2. As we only focused on one of these regions, the difference between the increase and decrease is beyond the scope of this work.

\subsection{Comparison with other works}
\par Our work suggests that the fraction of objects within our combined dataset which may be COs is at least 3.6\% in the most conservative case with, a significantly larger fraction than estimated by other works (e.g. 0.5\% \citep{2016ApJS..224...24L}.). This has potentially large ramifications for the confidence placed in photometric redshifts, particularly at the high end of the redshift regime. While our work does not correctly identify the same fraction of outliers we note these other works use different definitions of COs) as other ML methods, e.g. \citep{2022ApJ...928....6S} at roughly 80\% for some small number redshifts, more typically near 50\% based on their Figure 2 compared to our 55\%, our work achieves a false positive rate of 0. However, our work does match their 80\% true positive rate at a false positive threshold of only 10\% for our best network (and comparable values of the true positive rate and false positive rates in the other networks. In other works, the number of COs correctly identified, particularly in the area of interest for our study ($\approx \textrm{z-spec} > 2$), is a factor of 10 or greater less than the number of non outliers, making potential spectroscopic follow-up to confirm these results expensive with many false positives. Other works such as \citep{2021PASP..133d4504W, 2013MNRAS.431.2766S, 2008AJ....136.1361D, 2008ApJ...679...31M}, whose results are thoroughly summarised in \citep{2021PASP..133d4504W} use probability information to flag COs, but they also incorrectly flag a large number of NCOs to a degree likely similarly prohibitive to follow-up observations targeting COs. Again, we remind the reader these other works use different definitions of COs. Our definition is the most restrictive, focusing exclusively on the high spec-z, low photo-z range.

\section{Conclusions}
\label{sec:conclusions}
\par The primary result of this work is the construction of an ensemble of neural networks that accurately identifies catastrophic outliers with a true positive rate of 55.6\% for 8 band passes and 33.3\% for 5 band passes both with a false positive rate of 0\%. Additionally, these ensembles identified a list of 83,097 and 59,544 targets, respectively, which may also be previously unidentified COs within the COSMOS2020 dataset. Analysis of the networks suggests the most conservative estimate for the number of COs fitting our criteria is $\approx 3.6$\%. Lastly, when these are corrected with a toy regression model, it shifts the fraction of objects with photometric redshifts greater than 5 from 0.11\% to 1.6\% for the 8 band network and 0.24\% to 1.8\% for the 5 band network.
\par Developing methods to correct catastrophic outliers (COs) within large photometric catalogues is crucial. As there is not enough telescope time to provide spectroscopically determine redshifts, it is necessary to be able to develop methods that can provide both accurate photo-zs and also other methods that can independently determine when these photo-z algorithms are incorrect. Research conducted by \citep{2022ApJ...928....6S} demonstrates the feasibility of creating neural networks and other machine learning algorithms capable of accurately identifying a substantial fraction of catastrophic outliers within photometric redshift catalogues (note the difference in the definition of COs in their work). Our work takes this a step further by both developing a machine learning algorithm that correctly identifies a fraction of COs, while also doing it with few to no false positives. Because of the large imbalance between the NCOs and COs, a false positive rate of 0 or of an order where the false positives do not outnumber the true positives is helpful for identifying potential follow-up targets to validate the results of our classifier. Conducting additional spectroscopic and photometric follow-up observations on these COs will provide researchers with more samples to develop photo-z templates and/or ML algorithms that can accurately identify and correct COs in large galactic surveys.
\par Implementing data-driven corrections, as introduced in this work and others, can significantly improve photometric redshift estimates, thereby enhancing their reliability. Improving the reliability of photometric redshifts also has consequences for cosmology where photometric redshifts have been used to draw conclusions about the large scale structure of the universe. Furthermore, our work only applies a simple correction to the identified COs (fitting to a trend-line). As more of these types of objects are identified, more sophisticated methods could be developed (e.g. a custom SED template set) to improve these corrections further.
\par Future work on the COSMOS dataset could include the inverse of our designated outliers, (i.e. the objects with high reported phot-zs and low spec-zs). Another possible direction would be the incorporation of images of these objects. The reduction of the data to magnitudes could easily exclude useful information that may help the ML identify COs (e.g. nearby contaminants, compactness, etc).
\par Beyond the COSMOS dataset, future work could expand to other large datasets such as the North Ecliptic Plane (which has data in similar band passes) and other large photometric sky surveys. Furthermore, the CO problem is one example of the ``rare object detection problem" and a case study in unbalanced datasets. As modern instruments continue to collect thousands of images on millions of objects, there is a rapidly growing need for tools to pick out rare objects (e.g. metal poor and extremely metal poor galaxies) within these datasets (e.g. \citep{2005ASPC..347...66S}). While further work does need to be done to address the unbalanced dataset problem, our work shows it is possible to apply ML methods to find rare objects within these large astronomical datasets with a high degree of accuracy.

\section{Acknowledgements}
\label{sec:ack}
We gratefully acknowledge support for this research from NSF grant AST-1716093 (E.M.H., M.D.). We thank Amy Barger, Anthony Taylor, and Peter Sadowski for their helpful comments and suggestions.

\section{Software}
\texttt{NumPy} version 1.22.3 \citep{NumPy}, \texttt{Pandas} version 1.4.3 \citep{pandasDataStructure, pandasSoftware}, \texttt{Matplotlib} version 3.5.1 \citep{Matplotlib}, \texttt{Seaborn} version 0.12.0, \citep{Seaborn}, \texttt{Tensorflow} version 2.4.1 \citep{tensorflow2015-whitepaper}, \texttt{Keras} version 2.8.0 \cite{chollet2015keras}, \texttt{SciKit-Learn} version 1.0.2 \citep{Sklearn}, \texttt{imbalanced-learn} version 0.9.0 \citep{imblearn}

\bibliography{main}
\bibliographystyle{aasjournal}
\end{document}